\documentclass[preprint,showpacs,preprintnumbers,amsmath,amssymb,showkeys]{revtex4}

\usepackage{graphicx}
\usepackage{dcolumn}
\usepackage{bm}

\begin{document}

\preprint{Opt. Commun.}

\title{Anomalous wave propagation in quasiisotropic media}

\author{Hailu Luo}\email{hailuluo@gmail.com  (H. Luo).}
\author{Weixing Shu}
\author{Fei Li}
\author{Zhongzhou Ren}
\affiliation{ Department of Physics, Nanjing University, Nanjing
210008, China}
\date{\today}

\begin{abstract}
Based on boundary conditions and dispersion relations, the
anomalous propagation of waves incident from regular isotropic
media into quasiisotropic media is investigated. It is found that
the anomalous negative refraction, anomalous total reflection and
oblique total transmission can occur in the interface associated
with quasiisotropic media. The Brewster angles of E- and
H-polarized waves in quasiisotropic media are also discussed. It
is shown that the propagation properties of waves in
quasiisotropic media are significantly different from those in
isotropic and anisotropic media.
\end{abstract}

\pacs{78.20.Ci, 41.20.Jb, 42.25.Gy }
\keywords{Quasiisotropic medium; Negative refraction; Total
reflection; Oblique total transmission}
\maketitle

\section{Introduction}\label{Introduction}
In classic electrodynamics, it is well known that the
electrodynamic properties of anisotropic materials are
significantly different from those of isotropic materials. For
anisotropic materials, one or both of the permittivity
$\boldsymbol{\varepsilon}$ and the permeability $\boldsymbol{\mu}$
are second-rank tensors. The recent advent of a new class of
material with negative permittivity and permeability has attracted
much considerable
attention~\cite{Veselago1968,Smith2000,Shelby2001,Parazzoli2003,Houck2003,Mackay2004,Shen2006}.
Lindell {\it et al.} \cite{Lindell2001} have shown that anomalous
negative refraction can occur at an interface associated with an
anisotropic materials, which does not necessarily require that all
tensor elements of $\boldsymbol{\varepsilon}$ and
$\boldsymbol{\mu}$ have negative values. The studies of such
anisotropic materials have recently received much interest and
attention. Experimentally, an anisotropic metamaterial, which have
negative parameters of $\epsilon$ and $\mu$ tensors in the
microwave regime, has been constructed and
measured~\cite{Smith2003,Thomas2005}. Theoretically, anomalous
negative
refractions~\cite{Lindell2001,Belov2003,Smith2003,Thomas2005,Hu2002,Grzegorczyk2005a,Lakhtakia2004,Luo2005},
partial focus
lens~\cite{Smith2004a,Smith2004b,Parazzoli2004,Dumelow2005},
unusual quantum optical properties~\cite{Shen2004}, oblique total
transmission~\cite{Zhou2003} and  inverse Brewster
angle~\cite{Zhou2003,Grzegorczyk2005b} can be realized by
anisotropic metamaterials.

In general, E- and H- polarized waves propagate in different
directions in anisotropic media. Now a question arises: whether
there is a kind of anisotropic media in which  E- and H-polarized
waves propagate in the same direction? Recently, Shen {\it et al.
}~\cite{Shen2005} proposed a new kind of anisotropic media, which
was called quasiisotropic material. The quasiisotropic media have
some special properties: First, E- and H- polarized waves have the
same wave-vector surface. Second, in any allowed propagation
directions, E- and H- polarized waves propagate with the same
phase velocity and Poynting vector. Third, the wave vector and the
Poynting vector generally do not coincide with direction.

In this work, we present a detailed investigation on the
characteristics of electromagnetic wave propagation in
quasiisotropic media. We are interested in understanding the
condition under which the anisotropic media can be regarded as
quasiisotropic one. To obtain a better physical picture of the
total reflection and oblique total transmission, we introduce the
Brewster angles of E- and H-polarized waves in a quasiisotropic
media. We show that the propagation properties of waves in
quasiisotropic media are significantly different from those in
isotropic or anisotropic media.

\section{Dispersion relations of quasiisotropic media}\label{sec2}
In this section, we will present a brief investigation on the
refraction behavior at the interface between isotropic media and
quasiisotropic media. For anisotropic materials, one or both of
the permittivity and permeability are second-rank tensors. To
simplify the proceeding analyses, we assume the permittivity and
permeability tensors are simultaneously diagonalizable:
\begin{eqnarray}
\boldsymbol{\varepsilon}=\left(
\begin{array}{ccc}
\varepsilon_x  &0 &0 \\
0 & \varepsilon_y &0\\
0 &0 & \varepsilon_z
\end{array}
\right), ~~~\boldsymbol{\mu}=\left(
\begin{array}{ccc}
\mu_x &0 &0 \\
0 & \mu_y &0\\
0 &0 & \mu_z
\end{array}
\right).\label{matrix}
\end{eqnarray}
where $\varepsilon_i$ and $\mu_i$  are the permittivity and
permeability constants in the principal coordinate system
($i=x,y,z$). We consider the propagation of a planar wave of
frequency $\omega$ as ${\bf E}={\bf E}_0 e^{i {\bf k} \cdot {\bf
r}-i \omega t}$ and ${\bf H} = {\bf H}_0 e^{i {\bf k} \cdot {\bf
r}-i\omega t}$, through an isotropic media toward a quasiisotropic
LHM. In the principal coordinate system, Maxwell's equations yield
a scalar wave equation. In isotropic media, the accompanying
dispersion relation has the familiar form
\begin{equation}
 k_{x}^2+ k_{y}^2+k_{z}^2= \varepsilon_I \mu_I\frac{\omega^2}{c^2}, \label{D1}
\end{equation}
where $k_i$ is the $i$ component of the propagating wave vector,
$\omega$ is the frequency and $c$ is the speed of light in vacuum.
In the quasiisotropic medium, the E- and H- polarized incident
waves have the same dispersion relation~\cite{Shen2005}
\begin{equation}
\frac{ q_{x}^2}{\varepsilon_z \mu_y}+\frac{q_{y}^2}{\varepsilon_z
\mu_x}+\frac{q_{z}^2}{\varepsilon_y \mu_x}= \frac{\omega^2}{c^2},
\label{D2}
\end{equation}
where $q_{i}$ represents the $i$ component of transmitted
wave-vector.  For quasiisotropic media the permittivity and
permeability constants satisfy the quasiisotropy condition
\begin{equation}
\frac{\varepsilon_x }{\mu_x}=\frac{\varepsilon_y
}{\mu_y}=\frac{\varepsilon_z }{\mu_z}=C~~~(C>0), \label{QC}
\end{equation}
where $C$ is a constant. Based on the dispersion relation one can
find E- and H- polarized waves have the same wave-vector surface.
If $C>0$ the dispersion surface has the following two types:
ellipsoid and double-sheeted hyperboloid dispersion relations.
While $C<0$ the dispersion surface is single-sheeted hyperboloid.
It should be noted, as we will see the following, that the media
with single-sheeted hyperboloid dispersion relation can not be
regarded as quasiisotropic one.

We choose the $z$ axis to be normal to the interface, the $x$ and
$y$ axes  locate at the plane of the interface. The $z$-component
of the wave vector can be found by the solution of Eq. (\ref{D2}),
which yields
\begin{equation}
 q_z = \sigma\sqrt {\varepsilon_y \mu_x\frac{\omega^2}{c^2}-\frac{\varepsilon_y }{\varepsilon_z
}\left(\frac{\mu_x }{ \mu_y}q_{x}^2+q_{y}^2\right)}, \label{qz}
\end{equation}
where $\sigma=+1$ or $\sigma=-1$. The choice of $\sigma$ ensures
that light power propagates away from the surface to the $+z$
direction. In principle, the occurrence of refraction requires
that the $z$ component of the refracted wave vectors of  is real,
hence the incident angle must satisfy the following inequality:
\begin{equation}
\frac{\varepsilon_y }{\varepsilon_z }\left(\frac{\mu_x }{
\mu_y}q_{x}^2+q_{y}^2\right)<\varepsilon_y
\mu_x\frac{\omega^2}{c^2}.\label{IE}
\end{equation}
Without loss of generality, we assume the wave vector locate at
the $x-z$  plane ($k_y=q_y=0$). The incident angle is given by
\begin{equation}
\theta_I =\tan^{-1}\left[\frac{k_x}{k_{z}}\right]. \label{IA}
\end{equation}
The values of refractive wave vector can be determined by using
the boundary conditions and dispersion relations. The refractive
angle of the transmitted wave vectors of E- and H- polarized waves
can be written as
\begin{equation}
\beta_P^{E}=\tan^{-1}\left[\frac{q_x^{E}}{q_{z}^{E}}\right],~~~
\beta_P^{H}=\tan^{-1}\left[\frac{q_x^{H}}{q_{z}^{H}}\right].\label{PE}
\end{equation}

It should be noted that the actual direction of light is defined
by the time-averaged Poynting vector ${\bf S} =\frac{1}{2} {\bf
Re}({\bf E}^\ast\times \bf{H})$. For E- and H-polarized refracted
waves, ${\bf S}_T$ is given by
\begin{equation}
{\bf S}_T^{E}=Re \left[\frac{T_E^2 E_0^2 q_x^{E}}{2 \omega
\mu_z}{\bf e}_x+\frac{T_E^2 E_0^2 q_z^{E}}{2\omega\mu_x}{\bf
e}_z\right],\label{SE}
\end{equation}
\begin{equation}
{\bf S}_T^{H}=Re \left[\frac{T_H^2 H_0^2 q_x^{H}}{2 \omega
\varepsilon_z}{\bf e}_x+\frac{T_H^2 H_0^2
q_z^{H}}{2\omega\varepsilon_x}{\bf e}_z\right],\label{SH}
\end{equation}
where $T_E$ and $T_H$  are the transmission coefficient for E- and
H- polarized waves, respectively. The refraction angles of
Poynting vectors of E- and H- polarized incident waves can be
obtained as
\begin{equation}
\beta_S^{E}=
\tan^{-1}\left[\frac{S_{Tx}^{E}}{S_{Tz}^{E}}\right],~~~
\beta_S^{H}=
\tan^{-1}\left[\frac{S_{Tx}^{H}}{S_{Tz}^{H}}\right],\label{AS}
\end{equation}
As shown in Eqs.~(\ref{SE}) and~(\ref{SH}), one can see that the
direction of $z$ component of the Poynting vector is determined by
$q_z^{E}/\mu_x$  for E-polarized wave and $q_z^{H}/\varepsilon_x$
for H-polarized wave, respectively. Energy conservation requires
that the $z$ component of the energy current density of the
refracted waves must propagate away from the interface, for
instance, $q_z^{E}/\mu_x>0$ and $q_z^{H}/\varepsilon_x>0$. Since
the quasiisotropic condition
$\varepsilon_x/\mu_x=\varepsilon_z/\mu_z>0$ and boundary condition
$q_x^{E}=q_x^{H}=k_x$, $q_z^{E}$ and $q_z^{H}$ should have the
same sign. From Eqs.~(\ref{PE})-(\ref{SH}) we have
\begin{equation}
\beta_P^{E}= \beta_P^{H},~~~\beta_S^{E}= \beta_S^{H}.
\end{equation}
It means that both the wave vector and the Poynting vector of E-
or H-polarized field propagate in the same direction. It is one of
the significant differences between the quasiisotropic media and
anisotropic one.

Next we want to enquire: why the media with single-sheeted
hyperboloid dispersion relation can not be regarded as
quasiisotropic? Evidently, if
$\varepsilon_x/\mu_x=\varepsilon_z/\mu_z=C<0$, the signs of
$q_z^{E}$ and $q_z^{H}$ should be opposite.  From
Eqs.~(\ref{PE})-(\ref{SH}) we can obtain
\begin{equation}
\beta_P^{E}= -\beta_P^{H},~~~\beta_S^{E}= -\beta_S^{H}.
\end{equation}
It means that E- and H-polarized waves exhibit opposite
propagating properties in this kind of anisotropic media. The
inherent physics is stem from the fact ${\bf S}^E\cdot{\bf q^E}$
and  ${\bf S}^H\cdot{\bf q^H}$ have the opposite sign for E- and
H-polarized waves. Considering the properties of the
quasiisotropic medium, we conclude that the media with
single-sheeted hyperboloid dispersion relation can not be regarded
as quasiisotropic medium.

\section{Reflection and transmission coefficients }\label{sec3}
In this section, we will discuss the reflection and transmission
in the quasiisotropic media. For compactness, we assume the
electric filed polarized along the $y$ axis. For E-polarized
incident waves, the incident and reflected fields can be written
as
\begin{equation}
{\bf E} _{I} = {\bf e}_y E_0\exp[i(k_x x+k_z z)],\label{EI}
\end{equation}
\begin{equation}
{\bf E} _{R} =R_E E_0 {\bf e}_y \exp[i(k_x x-k_z z)],\label{ER}
\end{equation}
where $R_E$ is the reflection coefficient. Matching the boundary
conditions for each wave-vector component at the plane $z=0$ gives
the propagation field in the form
\begin{equation}
{\bf E} _{T}  = T_E E_0 \exp[i(q_x^{E} x+  q_z^{E} z)].\label{ET}
\end{equation}
Based on the continuity of the tangential component of the
electric field yields the equation
\begin{equation}
1+R_E=T_E.\label{EB1}
\end{equation}
A second equation can be found when we require continuity of the
tangential components of the ${\bf H}$ fields, which  can be
obtained from Maxwell's curl equation
\begin{equation}
{\bf H} =i
\frac{c}{\omega}\boldsymbol{\mu}^{-1}\cdot\nabla\times{\bf E}.
\end{equation}
Equating the $x$  component of the H vectors corresponding to the
incident, reflected, and transmitted fields, we have
\begin{equation}
\mu_x k_z(1-R_E )=\mu_I T_E  q_z^{E} ,\label{BBX1}
\end{equation}
For E-polarized incident waves, combining Eq.~(\ref{BBX1}) with
Eq.~(\ref{EB1}), one can obtain the following expression for the
reflection and transmission coefficients
\begin{equation}
R_E=\frac{\mu_x k_z-\mu_I q_z^{E}}{\mu_x  k_z+ \mu_I
q_z^{E}},~~~T_E = \frac{2 \mu_x k_z}{\mu_x  k_z+\mu_I
q_z^{E}}.\label{RETE}
\end{equation}
For H-polarized incident waves, the reflection and transmission
coefficients can be obtained similarly as
\begin{equation}
R_H=\frac{\varepsilon_x k_z-\varepsilon_I q_z^{H}}{\varepsilon_x
k_z+\varepsilon_I q_z^{H}},~~~T_H = \frac{2 \varepsilon_x
k_z}{\varepsilon_x k_z+\varepsilon_I q_z^{H}}.\label{RHTH}
\end{equation}
A comparison of Eq.~(\ref{RETE}) and  Eq.~(\ref{RHTH}) shows that
if  $\varepsilon_I/\mu_I=C$, one can obtain an interesting
property: E- and H-polarized waves will exhibit the same
reflection and transmission, namely $R_E=R_H $ and $T_E=T_H$.

We next explore the reflectivity and transmissivity which can be
defined by
\begin{equation}
r=\frac{S_{Rz}}{S_{Iz}},~~~t=\frac{S_{Tz}}{S_{Iz}},
\end{equation}
where $S_{Iz}$ and $S_{Tz}$ are the $z$ components of the incident
and transmission Poynting vectors,
respectively~\cite{Chen1983,Kong1990}. For E- and H- polarized
incident waves, the reflectivity and transmissivity can be written
as
\begin{equation}
r_E = R_E^2,~~~ t_E = \frac{\varepsilon_I q_z^{E}}{\varepsilon_x
k_z}T_E^2,\label{rE}
\end{equation}
\begin{equation}
r_H = R_H^2,~~~ t_H = \frac{\mu_I  q_z^{H}}{\mu_x
k_z}T_H^2.\label{rH}
\end{equation}
By using Eqs.~(\ref{rE}) and (\ref{rH}), one has $r+t=1$. This
demonstrates power conservation for reflection and transmission is
satisfied at the boundary surface.

Total reflection means the reflection coefficient is equal to
unity. Setting $R_E=1$ and $R_H=1$ in Eq.~(\ref{RETE}) and
Eq.~(\ref{RHTH}), one can obtain the critical angles as
\begin{equation}
\theta_C^{E}=\theta_C^{H}=\sin^{-1}\left[\sqrt{\frac{\varepsilon_z
\mu_y}{\varepsilon_I \mu_I}}\right],\label{CA}
\end{equation}
where $\varepsilon_z \mu_y>0$. If $\varepsilon_z \mu_y<0$ the
total reflection phenomenon never occur for any incident waves.

Total transmission  takes place at an incidence angle satisfying
$\theta_I+\beta_P^{H}=\pi/2$  for H-polarized wave incident from
free space into an isotropic dielectric RHM ($\mu_I=1$). Such an
angle, determined by
$\theta_B^{H}=\tan^{-1}[\sqrt{\varepsilon_I}]$, is called the
Brewster angle~\cite{Born1999,Jackson1998,Landau1984}. Similarly a
simple extension shows that the Brewster angles of E- and
H-polarized waves can be obtained in quasiisotropic media.
Mathematically the Brewster angles can be obtained from $r_E=0$
and $r_H=0$. For E-polarized incident wave if the anisotropic
parameters satisfy the relation
\begin{equation}
0<\frac{ \mu_z(\varepsilon_y \mu_I - \varepsilon_I
\mu_x)}{\varepsilon_I (\mu_I^2 -\mu_x \mu_z ) }< 1,\label{EBC}
\end{equation}
the Brewster angle can be expressed as
\begin{equation}
\theta_B^{E}=\sin^{-1}\left[\sqrt{\frac{ \mu_z(\varepsilon_y \mu_I
- \varepsilon_I \mu_x)}{\varepsilon_I (\mu_I^2 -\mu_x \mu_z ) }}
\right].\label{EB}
\end{equation}
For H-polarized wave if the anisotropic parameters satisfy by the
relation
\begin{equation}
0<\frac{ \varepsilon_z( \varepsilon_I \mu_y - \varepsilon_x\mu_I
)}{\mu_I (\varepsilon_I^2- \varepsilon_x \varepsilon_z)}<
1,\label{HBC}
\end{equation}
the Brewster angle can be written in the form
\begin{equation}
\theta_B^{H}=\sin^{-1}\left[\sqrt{\frac{ \varepsilon_z(
\varepsilon_I \mu_y - \varepsilon_x\mu_I  )}{\mu_I
(\varepsilon_I^2- \varepsilon_x \varepsilon_z) }}
\right].\label{HB}
\end{equation}
It should be mentioned that E- and H-polarized waves may exhibit a
Brewster angle simultaneously, which depends on the choice of the
anisotropic parameters in respect to the incident medium. The
Brewster angle can not exist for a certain material parameters
which make the corresponding expressions inside the square root
negative.

\section{Negative refraction and backward wave propagation }\label{sec3}
In this section, we will discuss the negative refraction and
backward wave propagation in quasiisotropic media. The refracted
wave in quasiisotropic media can be determined by the two
principles: First, the boundary conditions require that the
tangential component of the wave vector, is conserved across the
interface, $q_x=k_x$. Second, causality requires that the energy
current of the refracted waves should transmit away from the
interface, i.e., the normal component of the Poynting vector,
$S_{Tz}>0$. It is worth mentioning that the same principles are
also used to predict the refracted wave in more complex systems
such as multilayers~\cite{Russell1986,Zengerle1987}  and photonic
crystals~\cite{Luo2002,Luo2003,Foteinopoulou2003,Foteinopoulou2005,Yu2004,Jiang2005}.

First we want to study the anomalous negative refraction in
quasiisotropic media. Unlike in isotropic media, the Poynting
vector in quasiisotropic media is neither parallel nor
antiparallel to the wave vector, but rather makes either an acute
or an obtuse angle with respect to the wave vector. In general, to
distinguish the positive and negative refraction in quasiisotropic
media, we must calculate the direction of the Poynting vector with
respect to the wave vector. Positive refraction means ${\bf
q}_x\cdot{\bf S}_{T}>0$, and anomalous negative refraction means
${\bf q}_x\cdot{\bf S}_{T}<0$~\cite{Lindell2001,Belov2003}. From
Eqs.~(\ref{SE}) and (\ref{SH}) we get
\begin{equation}
{\bf q}_x\cdot{\bf S}_{T}^{E}=\frac{T_E^2 E_0^2 q_x^2}{2 \omega
\mu_z},~~~ {\bf q}_x\cdot{\bf S}_{T}^{H}=\frac{T_H^2 H_0^2
q_x^2}{2 \omega \varepsilon_z}.
\end{equation}
This anomalous negative refraction phenomenon is one of the most
interesting peculiar properties of quasiisotropic media. We can
see that the refracted waves will be determined by $\mu_z$ for
E-polarized incident waves and $\varepsilon_z$  for H-polarized
incident waves. The condition of negative refraction for both E-
and H-polarized waves can be given as
\begin{equation}
\varepsilon_z<0,~~~\mu_z<0.\label{NC}
\end{equation}
and other elements of both $\boldsymbol{\varepsilon} $ and
$\boldsymbol{\mu}$ do not need to be negative.

Next we want to investigate the backward wave propagation in the
quasiisotropic medium.  The wave with ${\bf q}\cdot{\bf S}_{T}<0$
has been called the backward wave or left-handed
wave~\cite{Veselago1968,Lindell2001,Belov2003}. From
Eqs.~(\ref{SE}) and (\ref{SH}) we have
\begin{equation}
{\bf q}^E \cdot{\bf S}_{T}^{E}=Re \left[\frac{T_E^2 E_0^2 q_x^2}{2
\omega \mu_z}+\frac{T_E^2 E_0^2
(q_z^E)^2}{2\omega\mu_x}\right].\label{SEE}
\end{equation}
\begin{equation}
{\bf q}^H \cdot{\bf S}_{T}^{H}=Re \left[\frac{T_H^2 H_0^2 q_x^2}{2
\omega \varepsilon_z}+\frac{T_H^2 H_0^2
(q_z^H)^2}{2\omega\varepsilon_x}\right],\label{SHH}
\end{equation}
Note that the Poynting vector and the wave vector generally are
not parallel or antiparallel in the quasiisotropic media. The
electric field  ${\bf E}$, the magnetic field  ${\bf H}$ and the
wave vector ${\bf q}$ can not form a strictly left-handed triplet.
Combining Eqs.~(\ref{D2}), (\ref{SEE}) and (\ref{SHH}), we can
find that
\begin{equation}
{\bf q}^E \cdot{\bf S}_{T}^{E}= \frac{T_E^2 E_0^2 \omega }{2
c^2}\varepsilon_y,~~~ {\bf q}^H \cdot{\bf S}_{T}^{H}= \frac{T_H^2
H_0^2 \omega }{2 c^2}\mu_y.
\end{equation}
Thus the conditions of backward wave propagation for both E- and
H-polarized waves can be written as
\begin{equation}
\varepsilon_y<0,~~~\mu_y<0,\label{BC}
\end{equation}
and other elements of both $\boldsymbol{\varepsilon} $ and
$\boldsymbol{\mu}$ need not to be negative. For the quasiisotropic
medium ($C>0$), in which $\varepsilon_y<0$ means automatically
$\mu_y<0$, so both E- and H-polarized waves can exhibit backward
wave propagation simultaneously.

Comparing Eq.~(\ref{NC}) with Eq.~(\ref{BC}) shows that the sign
of $ {\bf q}_x\cdot{\bf S}_T $ may not coincide with the sign of
${\bf q}\cdot{\bf S}_T$. Thus the negative refraction is not
necessarily tight to the backward wave propagation in
quasiisotropic media. Same has been found to be ture in uniaxially
anisotropic media~\cite{Belov2003}, and photonic
crystal~\cite{Foteinopoulou2003,Foteinopoulou2005}.

\section{Numerical results in quasiisotropic media}\label{sec3}
In this section, we will give the numerical results of waves
propagation in quasiisotropic media. The dispersion surface for
quasiisotropic media have the following two types: ellipsoid and
double-sheeted hyperboloid. In following two subsections we will
discuss the anomalous negative refraction, anomalous total
reflection and oblique total transmission in the two types of
media.

For the purpose of illustration, frequency contour will be used to
determine the refracted waves as shown in Fig.~\ref{Fig1} and
Fig.~\ref{Fig3}. From the boundary condition $q_x=k_x$, we can
obtain two possibilities for the refracted wave vector. Energy
conservation requires that the $z$ component of Poynting vector
must propagates away from the interface, for instance,
$q_z^{E}/\mu_x>0$ for E-polarized wave and
$q_z^{H}/\varepsilon_x>0$ for H-polarized wave. Then the sign of
$q_z$ can be determined easily. The corresponding Poynting vector
should be drawn perpendicularly to the dispersion contour. This is
because the Poynting vector is collinear with the ray
velocity~\cite{Born1999}. Depending on the sign of ${\bf q}\cdot
{\bf S}$, the Poynting vector is drawn inwards or outwards. For a
certain value of $k_x$ there are always two values of $q_z$ that
satisfy both the dispersion relation, while only the Poynting
vector with $S_{Tz}>0$ is causal.

\subsection{Ellipsoid dispersion relation}
\begin{figure}
\includegraphics[width=8cm]{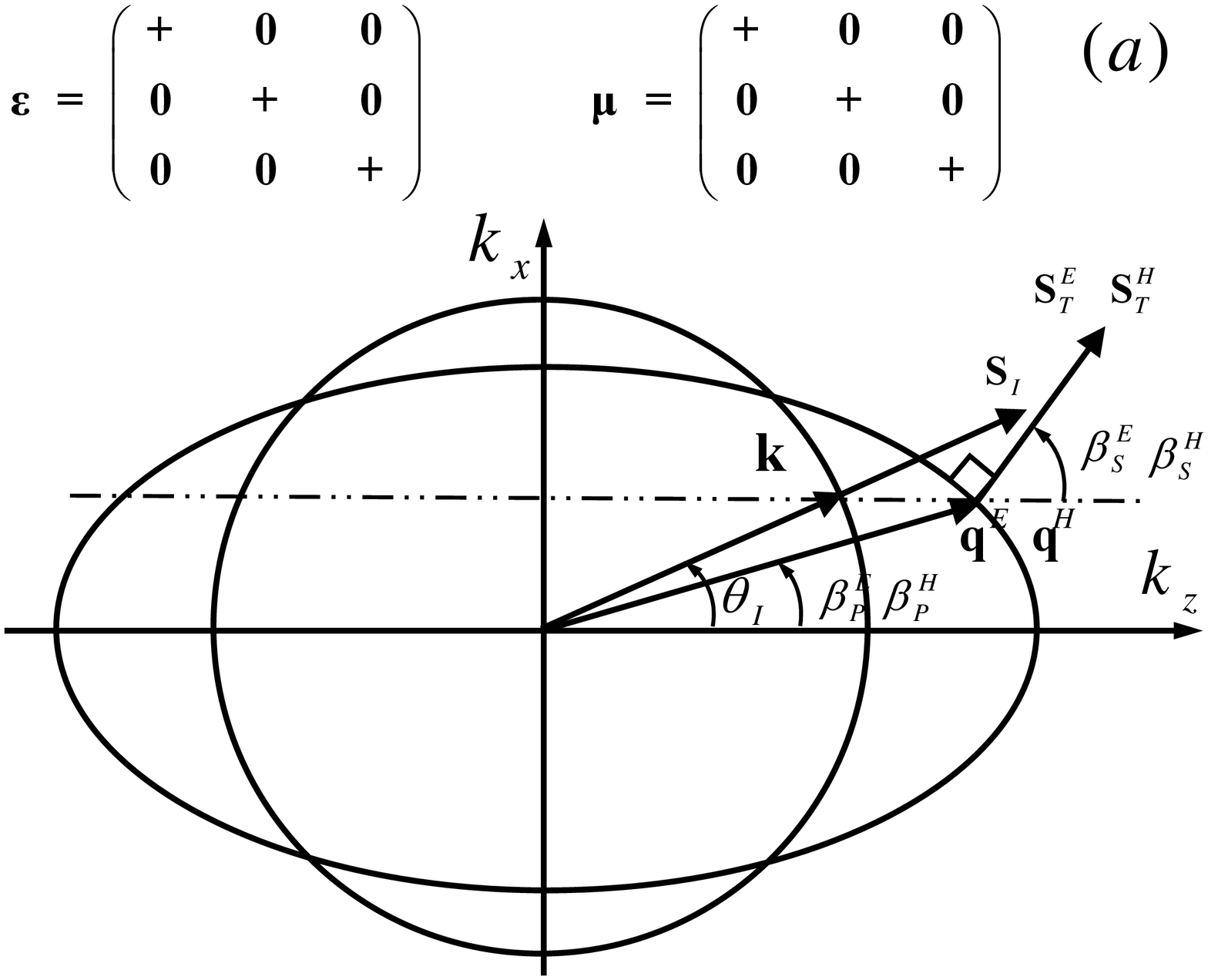}
\includegraphics[width=8cm]{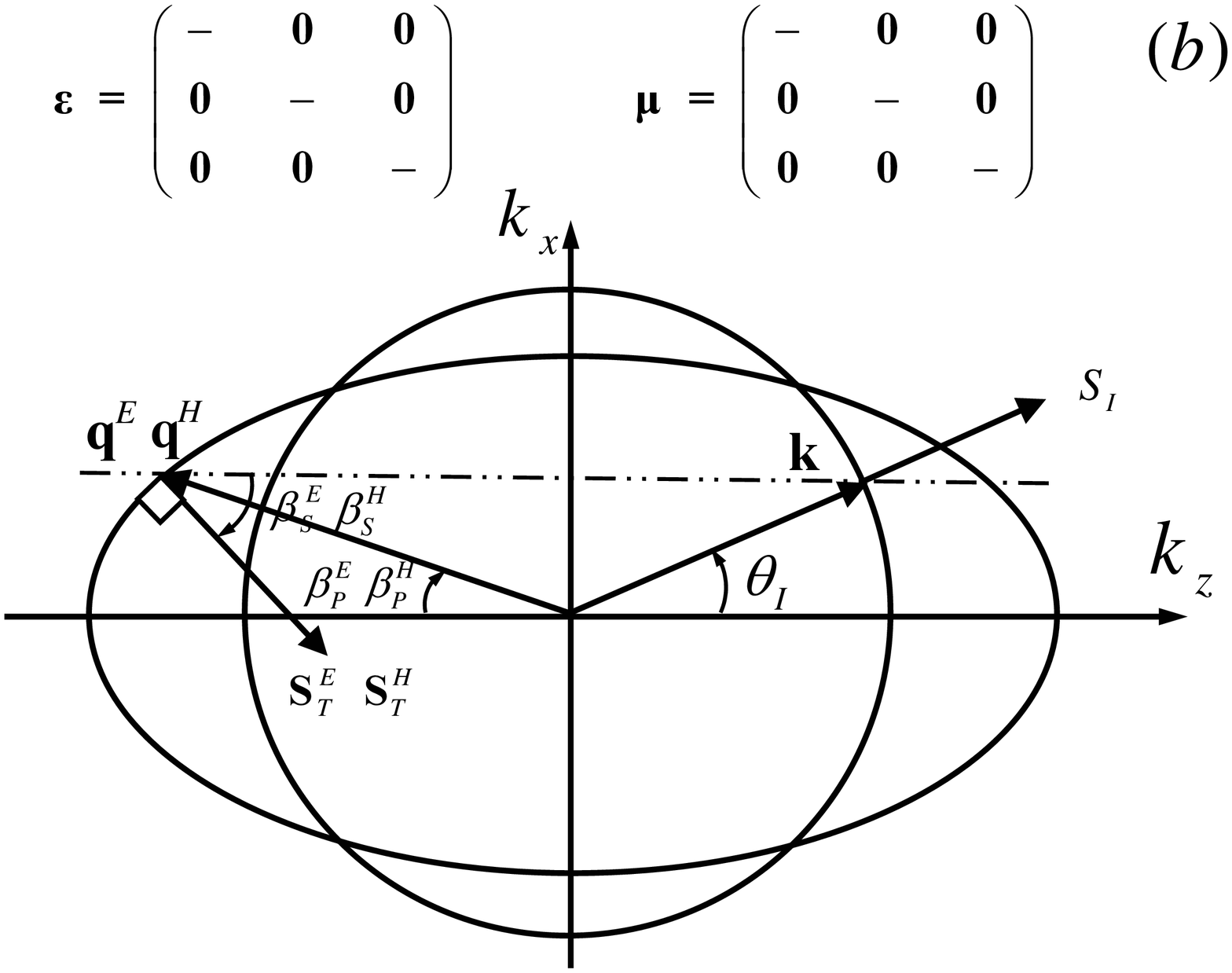}
\caption{\label{Fig1} The circle and the ellipse represent the
dispersion relations of isotropic media and quasiisotropic media,
respectively. The wave vectors and the Poynting vectors of E- and
H- polarized waves propagate in the same direction.  The ellipsoid
wave-vector surface has two subtypes: (a) corresponding to case I:
$\beta_P^{E}=\beta_P^{H}>0$, $\beta_S^{E}=\beta_S^{H}>0$. (b)
corresponding to case II: $\beta_P^{E}=\beta_P^{H}<0$,
$\beta_S^{E}=\beta_S^{H}<0$.}
\end{figure}

If all the $\varepsilon_i$ and $\mu_i$ have the same sign, the
parameter tensor elements wave-vector surface must be an ellipsoid
shown in Fig.~\ref{Fig1}. The ellipsoid dispersion relation has
two subtypes which can be formed from combinations of the material
parameter tensor elements. \\
Case I.~ For $\varepsilon_i>0$ and $\mu_i>0$, the refraction
diagram is shown in Fig.~\ref{Fig1}a. Here ${\bf k}_z\cdot{\bf
q}_{z}>0$ and ${\bf q}_x\cdot{\bf S}_{T}>0$, so the refraction
angles of wave-vector and Poynting vector are always positive. Due
to ${\bf q}\cdot{\bf S}_T>0$,
the backward wave can not propagate in this substype of media.\\
Case II. For $\varepsilon_i<0$ and $\mu_i<0$, the refraction
diagram is shown in Fig.~\ref{Fig1}b. Here ${\bf k}_z\cdot{\bf
q}_{z}<0$ and ${\bf q}_x\cdot{\bf S}_T<0$, so the refraction angle
of wave vector and Poynting vector are always negative. The
backward wave can propagate in this substype of media since ${\bf
q}\cdot{\bf S}_T<0$.

In principle, the occurrence of refraction requires that the $z$
component of the wave vector of the refracted waves must be real.
If $\varepsilon_z \mu_y<\varepsilon_I \mu_I$ the real wave vector
only exists for the branch
\begin{equation}
-\theta_{C}<\theta_{I}<\theta_{C}.
\end{equation}
When $\theta_I>\theta_C$ and $r_E=r_H=1$, the total reflection of
E- and H-polarized waves occur in the branch
$-\pi/2<\theta_{I}<-\theta_{C}$ and $\theta_{C}<\theta_{I}<\pi/2$.
For E- and H-polarized waves, $\varepsilon_z \mu_y$ is negative,
thus the inequality $\varepsilon_z \mu_y<\varepsilon_I \mu_I$
satisfied for any incident angle. The anomalous total reflection
phenomenon thus can not occur.

\begin{figure}
\includegraphics[width=12cm]{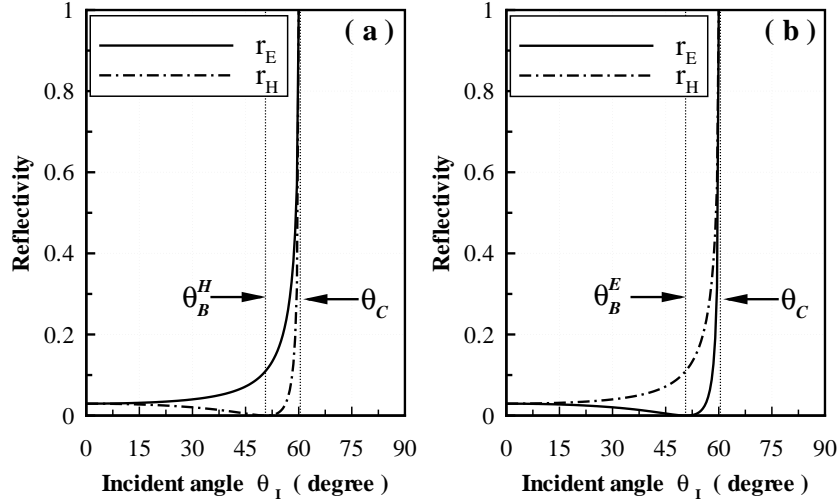}
\caption{\label{Fig2} The reflectivity of E- and H-polarized waves
as functions of the incident angle $\theta_I$. The quasiisotropic
medium with parameters $\varepsilon_x=\varepsilon_y=1$,
$\varepsilon_z=1.5$ and $C=1$. (a) H-polarized incident wave
exhibits a Brewster angle. The isotropic medium with
$\varepsilon_I=1$ and $\mu_I=2$. (b) E-polarized incident wave
exhibits a Brewster angle. The isotropic medium with
$\varepsilon_I=2$ and $\mu_I=1$. }
\end{figure}

In the next step, we will discuss the interesting phenomenon of
oblique total transmission $T_{(\theta_I\neq0)}=1$.  The
reflectivity of E- and H-polarized waves are plotted in
Fig.~\ref{Fig2}. We choose some simple parameters for the purpose
of illustration, i.e., $C=1$. In Fig.~\ref{Fig2}a H-polarized
incident wave exhibits a Brewster angle. When
$\theta_I=\theta_B^{H}$, one can find $r_H= 0$ and H-polarized
incident wave can exhibit oblique total transmission. In
Fig.~\ref{Fig2}b E-polarized wave exhibits a Brewster angle. When
$\theta_I=\theta_B^{E}$, one can find $r_E= 0$ and E-polarized
incident waves can exhibit oblique total transmission. The
appearance of the Brewster's angle for E-polarized waves is due to
the quasiisotropic medium having primarily a magnetic response
that reverses the roles of E- and H-polarization.

\subsection{Double-sheeted hyperboloid dispersion relation}
\begin{figure}
\includegraphics[width=8cm]{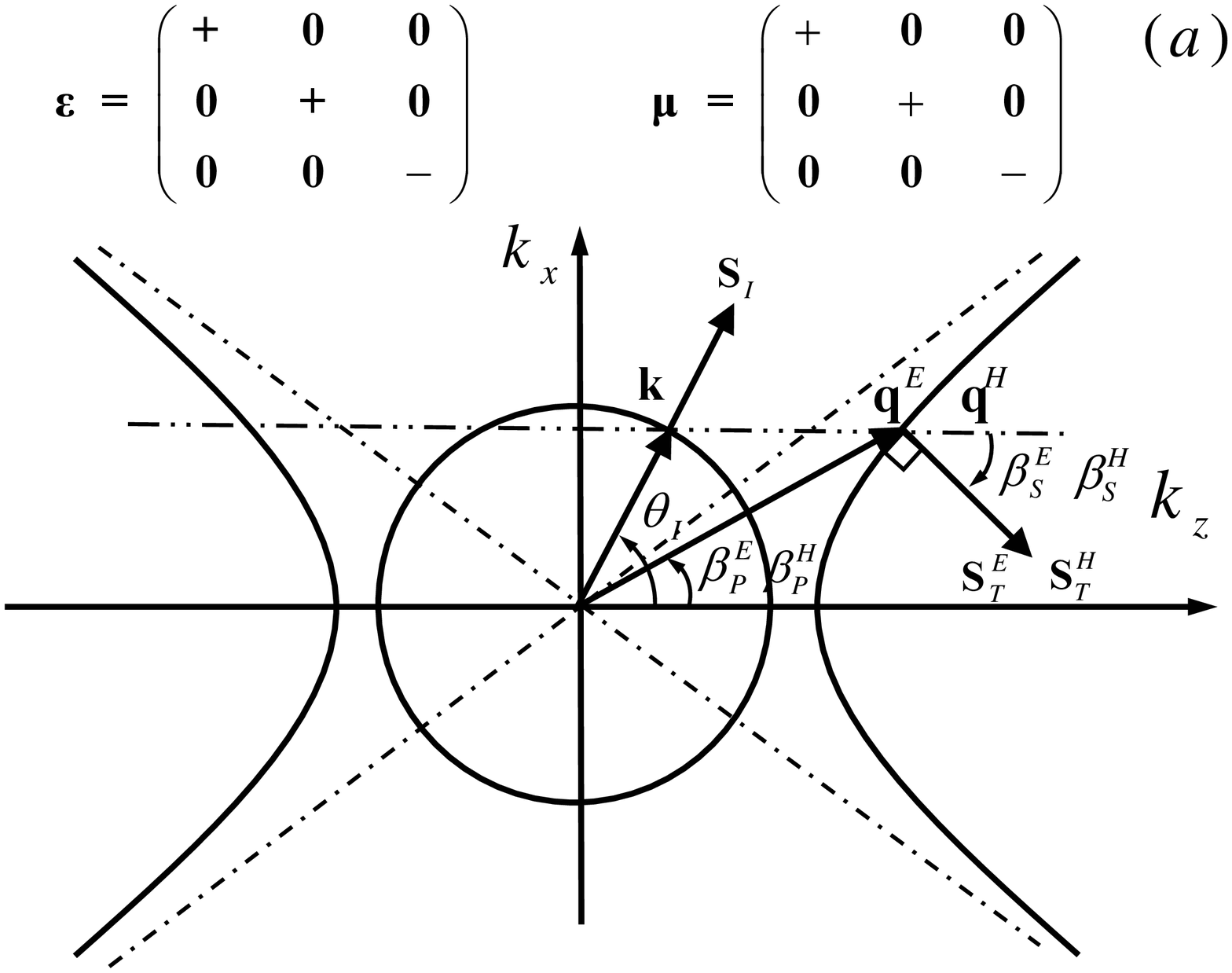}
\includegraphics[width=8cm]{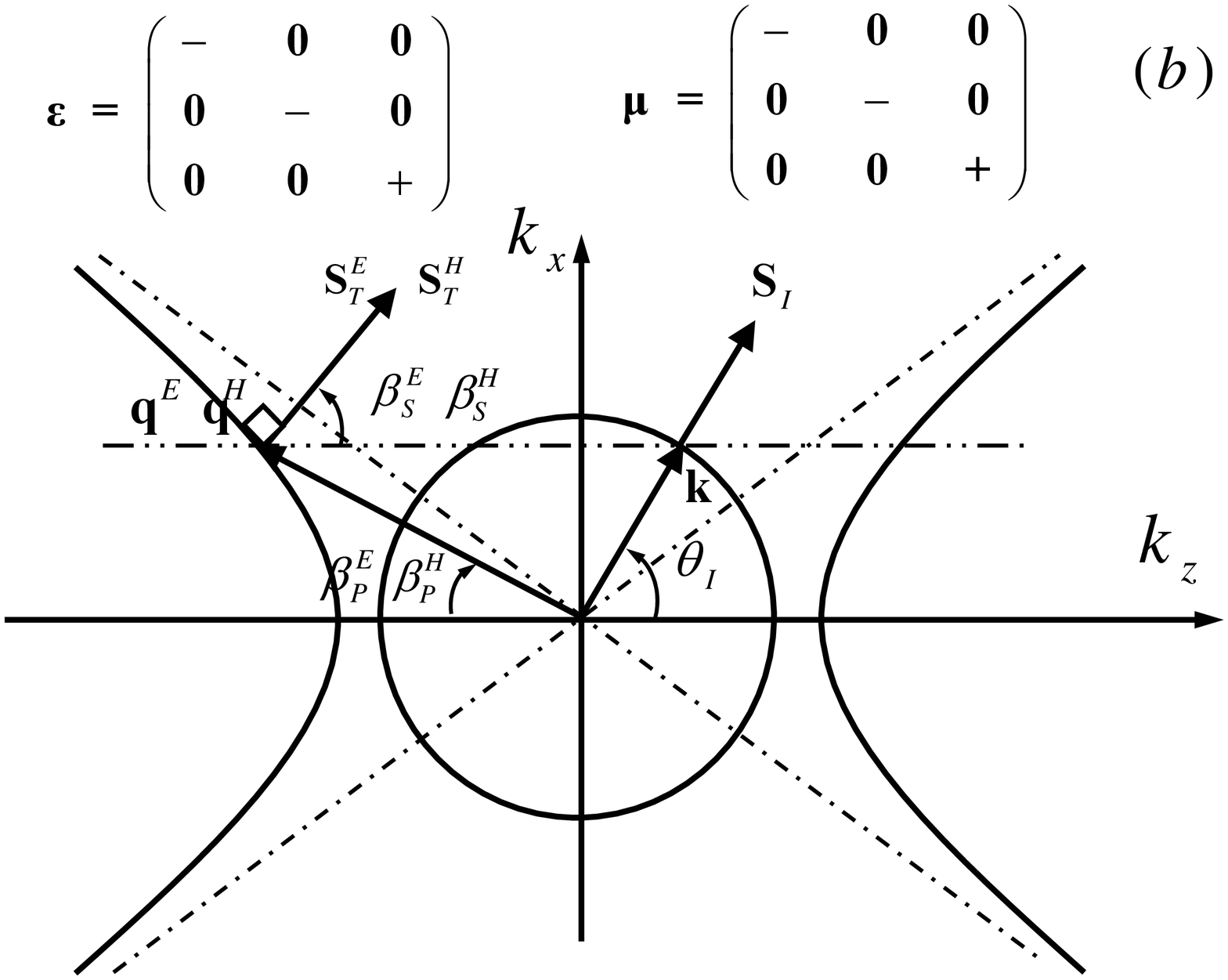}
\caption{\label{Fig3} The circle and the double-sheeted hyperbola
represent the dispersion relations of isotropic media and
quasiisotropic media, respectively. The wave vector and Poynting
vectors of E- and H- polarized waves propagate in the same
direction.  The double-sheeted hyperboloid wave-vector surface has
two subtypes: (a) corresponding to case I:
$\beta_P^{E}=\beta_P^{H}>0$, $\beta_S^{E}=\beta_S^{H}<0$. (b)
corresponding to case II: $\beta_P^{E}=\beta_P^{H}<0$,
$\beta_S^{E}=\beta_S^{H}>0$.}
\end{figure}

If only two of $\varepsilon_i$ have the same sign, for instance
$\varepsilon_x\cdot\varepsilon_y>0$, the wave-vector surface is a
double-sheeted hyperbola. Also this media
has two subtypes:\\
Case I.~ For the case of $\varepsilon_x>0$, $\varepsilon_y>0$ and
$\varepsilon_z<0$, the refraction diagram is shown in
Fig.~\ref{Fig3}a. Here ${\bf k}_z\cdot{\bf q}_{z}>0$ and ${\bf
q}_x\cdot{\bf S}_{T}<0$. It yields that the refraction of Poynting
vector refraction is always negative even if the wave-vector
refraction is always positive.
Due to ${\bf q}\cdot{\bf S}_T>0$, the backward wave can not propagate in this substype of media.\\
Case II. For the case of $\varepsilon_x<0$, $\varepsilon_y<0$ and
$\varepsilon_z>0$, the refraction diagram is shown in Fig.
~\ref{Fig3}b. Here ${\bf k}_z\cdot{\bf q}_{z}<0$ and ${\bf
q}_x\cdot{\bf S}_{T}>0$, the refraction of Poynting vector is
always positive, even if the refraction of wave vector refraction
is always negative. The backward wave can propagate in this
substype of media since ${\bf q}\cdot{\bf S}_T<0$.

It is interesting to observe that the anomalous backward waves can
propagate in this kind of media. Fig.~\ref{Fig3}a and
Fig.~\ref{Fig3}b have presented simple examples of possibilities
of negative refraction without backward waves, and backward waves
without negative refraction, respectively.

For E- and H-polarized waves, $\varepsilon_z \mu_y$ is negative,
thus the inequality $\varepsilon_z \mu_y<\varepsilon_I \mu_I$
satisfied for any incident angle. From Eq.~(\ref{IE}) the real
wave vector exists for the branch
\begin{equation}
 -\pi/2<\theta_{I}<\pi/2.
 \end{equation}
In this case, the anomalous total reflection phenomenon can not
occur for any incident angles.

\begin{figure}
\includegraphics[width=12cm]{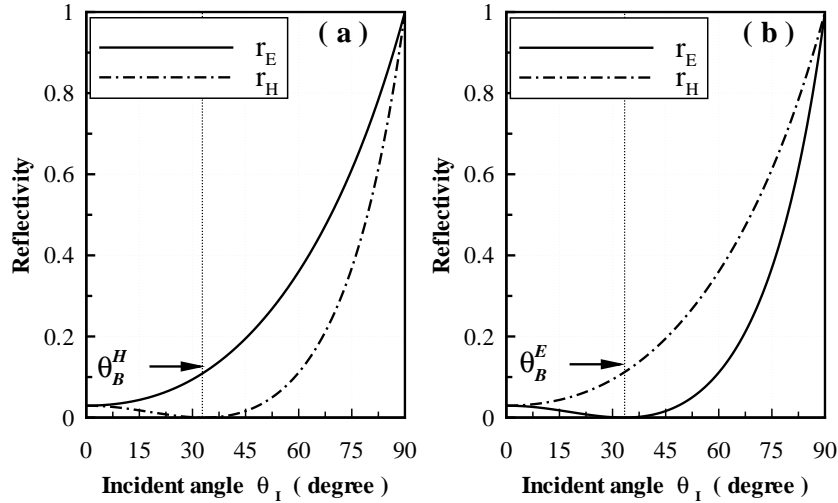}
\caption{\label{Fig4} The reflectivity of E- and H-polarized waves
as functions of the incident angle $\theta_I$. The quasiisotropic
medium with parameters $\varepsilon_x=\varepsilon_y=1$,
$\varepsilon_z=-1.5$ and $C=1$. (a) H-polarized incident wave
exhibits a Brewster angle. The isotropic medium with
$\varepsilon_I=1$ and $\mu_I=2$.  (b) E-polarized incident wave
exhibits a Brewster angle. The isotropic medium with
$\varepsilon_I=2$ and $\mu_I=1$.}
\end{figure}

The reflectivity of E- and H-polarized waves are plotted in
Fig.~\ref{Fig4}. As expected, H-polarized wave can exhibit a
Brewster angle as shown in Fig.~\ref{Fig4}a. When
$\theta_I=\theta_B^{H}$, one can find that $r_H= 0$ and
H-polarized incident waves can exhibit oblique total transmission.
Let us now consider the second case in which E-polarized wave
exhibits a Brewster angle, as depicted in Fig.~\ref{Fig4}b. When
$\theta_I=\theta_B^{E}$, one can find $r_E= 0$ and E-polarized
incident waves can exhibit a oblique total transmission. The
inherent physics underlying  the oblique total transmission are
collective contributions of the electric and magnetic responses.
The Brewster condition discussed here requires that the reflected
fields of the oscillating electric and magnetic dipoles cancel
each other for certain incident angle.

Finally we want to enquire whether  E- and H-polarized waves have
the same Brewster angle. Obviously, if $\varepsilon_I/\mu_I=C$,
Eqs.~(\ref{EBC}) and (\ref{HBC}) are satisfied simultaneously,
then E- and H-polarized waves present the same Brewster angle,
i.e., $\theta_B^{E}=\theta_B^{H}$.  For the two polarized waves
the material thus present oblique total transmission at the same
incident angle.

\section{Conclusion }\label{sec4}
In conclusion, we have investigated the properties of waves
propagation in quasiisotopic media. We have performed the detailed
analyses of the anomalous negative refraction, anomalous total
reflection and oblique total transmission at the interface
associated with quasiisotropic media. To obtain a better physical
picture of the total reflection and oblique total transmission, we
have introduced the Brewster angles for E- and H-polarized waves
in quasiisotropic media. We have shown that the properties of
waves propagation in quasiisotopic media are significantly
different from those in isotropic or anisotropic media. It should
be pointed out that the anisotropic media has been fabricated
successfully~\cite{Smith2003,Thomas2005}. We are justified to
expect that the quasiisotropic medium can be constructed quickly
and the anomalous propagation characteristics will be studied
experimentally. Finally, it is worth pointing out that the
polarization insensitive lens can be designed based on anomalous
propagation properties discussed above.

\begin{acknowledgements}
H. Luo is sincerely grateful to Professors W. Hu and Q. Guo for
many fruitful discussions. We also wish to thank the anonymous
referees for their valuable suggestions. This work was supported
by projects of the National Natural Science Foundation of China
(Nos. 10125521 and 10535010).
\end{acknowledgements}

\end{document}